\newcommand{\s}{\mbox{$\sigma _{1}$}}
\newcommand{\st}{\mbox{$\sigma _{2}$}}
\newcommand{\sth}{\mbox{$\sigma _{3}$}}
\newcommand{\NT }{\mbox{$N ^{T}$}}
\newcommand{\N }{\mbox{$N $}}
\documentstyle[12pt]{article}
\newcommand{\th}{\mbox {$ \tilde{\theta}$}}
\newcommand{\li}{\mbox {$    \lambda^{-1}_{n,m}$}}
\newcommand{\lp }{\mbox {$e^{i\int _{c} k(t) \partial _{z} X(z+t) dt}$}}
\newcommand{\lpp}{\mbox {$e^{i\int _{c} p(t) \partial _{z} X(z+t) dt}$}}
\newcommand{\bp }{\mbox {$e^{i\int d \sigma \int _{c} k(t, \sigma )
\partial _{z} X(z+t, \sigma )dt}$}}
\newcommand{\Pl}{\mbox {$ P^{\lambda}$}}
\newcommand{\Pm}{\mbox {$ P^{\mu}$}}
\newcommand{\Prh}{\mbox {$ P^{\rho}$}}
\newcommand{\la}{\mbox{$ \lambda $}}
\newcommand{\be}{\begin{equation}}
\newcommand{\br}{\begin{eqnarray}}
\newcommand{\ee}{\end{equation}}
\newcommand{\er}{\end{eqnarray}}
\newcommand{\gvq}{\mbox {$ e^{i\sum _{n\geq 0}q_{n}Y_{n}}$}}
\newcommand{\gvk}{\mbox {$ e^{i\sum _{n > 0}k_{n}Y_{n}}$}}

\newcommand{\eln}{\mbox {$ e^{\sum _{n}\lambda _{n}L_{-n}}$}}
\newcommand{\emn}{\mbox {$ e^{\sum _{n}\mu _{n}L_{-n}}$}}
\newcommand{\ern}{\mbox {$ e^{\sum _{n}\rho _{n}L_{-n}}$}}
\newcommand{\tli}{\mbox {$( \theta - \la ^{-1})^{-1}$}}
\newcommand{\tmi}{\mbox {$( \theta - \mu ^{-1})^{-1}$}}
\newcommand{\tl}{\mbox {$( \theta - \la ^{-1})$}}
\newcommand{\tm}{\mbox {$( \theta - \mu ^{-1})$}}
\renewcommand{\theequation}{\thesection.\arabic{equation}}
\begin{document}
\title{Loop Variables and the Virasoro Group}
\author{B. Sathiapalan\\ {\em
Physics Department}\\{\em Penn State University}\\{\em 120
Ridge View Drive}\\{\em Dunmore, PA 18512}}
\maketitle
\begin{abstract}
We derive an expression in closed form for the action of a finite
element of the Virasoro Group on generalized vertex operators.  This
complements earlier results giving an algorithm to compute the action
of a finite string of generators of the Virasoro Algebra on generalized
vertex operators.  The main new idea is to use a first order formalism
to represent the infinitesimal group element as a loop variable.  To
obtain a finite group element it is necessary to thicken the loop to a
band of finite thickness.  This technique makes the calculation very
simple.
\end{abstract}
\newpage
\section{Introduction}

Understanding the Virasoro group is of central importance in
string theory
since the gauge invariance of the theory is a
consequence of the conformal invariance of the two dimensional
world sheet field theory and the generators of infinitesimal conformal
transformations obey the Virasoro Algebra.
A lot of work has been done on the
representation theory of the Virasoro Algebra and applications to various
branches of physics\cite{II,III,IV,V,VI,VII,VIII,IX,X}.  For many
purposes it is sufficient to consider
the algebra
rather than the group.  In particular in string theory attention has
been focussed on infinitesimal gauge transformations and for
this the algebra is
 all one needs to consider.
 Experience with Yang-Mills theories and gravity teaches
us however that it is essential to consider the group in order to
appreciate
the geometrical basis underlying gauge symmetries.  Another motivation for
studying the Virasoro group is that the string evolution operator
  $e^{L_{0}\tau }$ is obviously an element of the Virasoro group.  It is
plausible that a gauge invariant generalization of this would be
$e^{\sum_{n}L_{n}\tau_{n}}$ which is a general element of the Virasoro
group.
    The Virasoro group also has been studied from different points of
view \cite{XI,XII,XIII,XIV,XV,XVI}.
Our goal in this paper is to approach these issues using the loop
variable formalism.  This will also improve our understanding of the
formalism which we feel will be useful for developing computational
techniques for string theory.  We derive an expression
for the action of a Virasoro group element on generalized vertex
operators,i.e.
we write down an expression for
$e^{\sum_{n} \lambda _{n} L_{-n}}V(z)\mid > $
where $V(z) $is a loop variable or generalized vertex operator (gvo).
\begin{equation}
e^{i(k_{0}X + \int dt k(t) \partial _{z}X(z+t))} =
 e^{i\sum _{n}\frac{k_{n}\partial ^{n} X(z)}{(n-1)!}} \equiv
 e^{ik_{n}Y_{n}}
\end{equation}

(A sum over the index $n$ is understood.)

A loop variable contains in it all the vertex operators in the
bosonic string theory \cite{XVII,XVIII}.  The Taylor expansion in
$t$ is allowed
 because
there are no singularities in $t$.  This will not be the case if we
consider correlation functions with other operators located inside the
contour.  Thus in general we must make a distinction between the LHS
of (1.1), which is a loop variable and the RHS, which is a generalized
vertex operator. The loop variable is more general than the generalized
vertex operator.  In \cite{XVIII} we described a simple algorithm for
calculating the action
of an arbitrary finite sequence of $L_{n}$'s on the loop variable.  Here
instead of an {\em algorithm} we have a mathematical {\em formula}
involving some definite matrices and their products for the action of
a {\em finite group element}.  The formula is as
follows:
We define
\begin{equation}
\la _{n,m} ( \sigma ) = \la _{m,n} ( \sigma ) = \la _{n+m} \ : 0 \leq
\sigma  \leq 1 ; n,m \in Z
\end{equation}

\begin{equation}
\th _{ \s , \st } =   \N \theta ( \s - \st ) + \NT \theta ( \st - \s ) ;
\theta (x) = \begin{array}{cc} 1 &  x>0 \\ 0 & x<0  \end{array}
\end{equation}

where $N$ is an infinite dimensional matrix acting on infinite dimensional
vectors of the form $y_{n},n \in Z$.  If we consider the subspace
$\left( \begin{array}{l}y_{n} \\ y_{-n} \end{array} \right) ,n>0$
it has the form
\(
\N = \left( \begin{array}{cc} 0 & 0 \\ n & 0 \end{array} \right)
\) .
When $n=0$,$N= \left( \begin{array}{cc}0 & 0 \\ 1 & 0 \end{array} \right)$.
We make a distinction between $y_{+0}$ and $y_{-0}$ and set the former
equal to zero i.e.$y_{+0}=0$.  Finally, define the (infinite) column vector
${\bf Y} = \left( \begin{array}{l} Y_{n} \\ Y_{-n} \end{array} \right)$
where $ Y_{-n} \begin{array}{lll} \equiv & -ink_{n} & n>0 \\
\equiv & -ik_{0} & n=0
\end{array}  $
and $Y_{+n} \equiv Y_{n} \equiv \partial ^{n}X/ (n-1)! , n>0$
 and $Y_{+n}\equiv 0 ,n =0 $

In terms of these variables:
\be
\eln : \gvq : = : e^{ -1/2 {\bf Y}^{T} \Pl {\bf Y}} \gvq :
Det [1 - \la \th ]
\ee
with
\be
\Pl = \int d \s \left\{ \frac {1}{ \delta _{ \s \st } - \la ( \s ) \th
 _{ \s \st } } \right\} \la ( \st )
\ee
The object in curly brackets is a matrix with continuous indices
\s\ ,\st\
and multiplies a column vector \la (\st ).
Thus \st\ is summed over.  Inside the
curly bracket there is no sum
on \s\ .\Pl\ can be represented as an infinite series in
\la\
  by expanding the expression in curly brackets.
\begin{eqnarray}
\Pl\ & = & \int_{0}^{1} d\s\ \la (\s )
+ \int_{0}^{1} d\s\ \int_{0}^{1} d\st\ \la (\s ) \th (\s\ , \st )
 \la (\st )+ \nonumber \\ &  &
 \int_{0}^{1} d\s\ \int_{0}^{1} d\st\ \int_{0}^{1} d\sth\
\la (\s ) \th (\s\ , \st ) \la (\st) \th (\st\ ,\sth ) \la ( \sth )
+... \nonumber \\
     & = & \la\ + 1/2 \la (N+N^{T}) \la  + 1/3![ \la N\la N \la +
\la N^{T}\la N^{T} \la\ + \nonumber \\  &  &
 2\la N \la N^{T} \la\ + 2\la N^{T} \la N \la ] +...
\end{eqnarray}
Thus $e^{1/2 {\bf Y}^{T} \Pl {\bf Y}}$ can be regarded as a representation
of the group element $e^{\lambda _{n} L_{-n}}$.  There is a well defined
group composition law$\Pl \circ  \Pm =  P ^{\rho ( \lambda ,\mu  )}$.
 When
\la\ and $\mu$ are infinitesimal this law satisfies the Virasoro Algebra.
    Equations (1.4,1.5) summarize the main result of this paper.  They
describe the action of an element of the Virasoro group on
an arbitrary vertex operator.

    This paper is organized as follows: In section II we motivate the
result
in an intuitive way. In section III we give a more complete proof.
In section
IV we discuss the group composition property.  This can be used to
construct an alternate proof of the result. A sample calculation
is also given.
In section V we give some concluding remarks.
\newpage
\section{Loop Variable Representation of Group Element}
\setcounter{equation}{0}
    Consider the integral
\begin{equation}
\int {\cal D}k(t) \lp\  \Phi [k(t)]
\end{equation}

with $k(t) = k_{0} + \frac{k_{1}}{t} + \frac{k_{2}}{t^{2}} +...$ which
is a loop variable representation of (D-dimensional) fields of the bosonic
string as considered in \cite{XVII}.
(Later we will generalize to include
positive powers of $t$ also).
If we choose
\begin{equation}
\Phi [k(t)] = e^{-1/2 \int dt k^{2} (t) \lambda ^{-1} (t)}
\end{equation}
and perform the functional integral over $k(t)$ we get, upto an overall
normalization factor :
\be
e^{-1/2\int dt \partial X(z+t) \partial X(z+t) \lambda (t)}
[Det \la ]^{D/2}
\ee
The exponent is nothing but the energy momentum tensor multiplied by a
gauge parameter.  If we choose
\be
\la (t) = \la _{0} t + \la _{1} + \la _{2}/t + ...+\la _{-1} t^{2} +
\la _{-2} t^{3} +...   + \la _{n} t^{-n+1} +...
\ee
then (2.3) becomes $e^{\sum _{n} \lambda _{n} L_{-n}}$which is an
element of the
Virasoro group.  The $\la_{n}$in (2.4) is the same as in (1.2).
The above
discussion has to be modified to include the effects of normal ordering
when we make $X$ a two dimensional quantum field.  Furthermore $k(t)$ will
also include positive powers of $t$ when we consider operator
products of the loop variable (2.1) with other vertex operators
at the point $z$.  Nevertheless
the basic idea illustrated above will be used.  Namely we will use the
first order representation (2.1) instead of (2.3).
In order to calculate the action of a group element on a vertex
operator we first perform the functional
integral over $X$ to get the terms in the operator product expansion of
(2.1) with the vertex operator and then do the $k(t)$ integration.

    Now we have to incorporate quantum mechanics since otherwise the
$L_{n}$'s in (2.3) will commute.  That means we have to include normal
ordering effects.  We can try the following:  Write
\be
\lp\  = : \lp\ : e^{1/2 \int dt k(t) < \partial X(z+t)
\frac{\partial ^{n} X(z)}{(n-1)!} > k_{n}}
\ee
In order to avoid the ambiguity at $t=t'$ we
will take one loop to have a slightly larger radius as shown in Fig.1.
On the inner loop $X(z+t)$ can be expanded in a Taylor series in $t$
(i.e. in positive powers of $t$) since there is no singularity as the
radius of the loop shrinks to zero. We have used
$<X(z) X(w)>= ln(z-w)$ to get for the second factor
\be
e^{\sum _{m>0} m k_{m}.k_{-m}}
\ee
where
\be
k(t) = \sum _{m =-\infty}^{+\infty}k_{m}t^{-m}
\ee
Thus (2.5) becomes
\be
: \lp\ :  e^{\sum _{-\infty}^{+\infty} 1/2 k_{n}.k_{m} \theta _{nm}}
\ee
where
\be
\theta _{n,m}  =   m\delta _{n,-m} ,m \neq 0
\ee
and (2.1) becomes
\be
\int {\cal D} k(t) : \lp\ :
e^{1/2k_{n}[ \theta _{n,m} - \lambda ^{-1} _{n,m} ] k_{m}}
\ee
where \li\ is the inverse of $ \la _{n,m}$,the matrix defined in (1.2).
It satisfies
$\li = (\la ^{-1}) _{n+m}$
 where $\la ^{-1}(t) = (\la^{-1}) _{n} t^{-n+1}$.  In fact it will turn
out that (2.9) is not the correct matrix to be used nor is (2.10) the
correct final answer.  But it is quite close.  Let us try to use (2.10) in
calculating the action of a Virasoro group element on a
vertex operator.  In the process we will see both the
facility of this approach as well as the ordering
subtleties that necessitate a further refinement of the steps
leading from
(2.5) to (2.10).
    We want to calculate the operator product of (2.10) with a
    generalized
vertex operator  :\gvq\ : where
\be
Y_{n}(z) = \frac{\partial ^{n}X}{(n-1)!}
 ; Y_{0} =X(z)
\ee
Thus we have to evaluate the operator product (OP):
\be
\int {\cal D}k(t) : \lp : : \gvq : e^{1/2k_{n}[\theta _{nm} -
\lambda ^{-1}_{nm}]k_{m}}
\ee
Using (we assume that the vertex operator is located at $z$)
\be
<\partial _{z} X(z+t) \partial ^{n}X(z)> = \frac{n!}{t^{n+1}}
\ee
we get for (2.12):
\be
\int {\cal D}k(t) :\lp \gvq : e^{\sum _{n>0}k_{-n}.q_{n}n +k_{0}.q_{0}}
e^{1/2k_{n}[ \theta _{nm} - \lambda ^{-1}_{nm}] k_{m}}
\ee
Singularities in $t$ arise only when (2.14) is inserted in correlators
with other vertex operators at $z$.  As long as we refrain from doing that
one can Taylor expand $\partial X(z+t) $ in (positive) powers of $t$
 to get
\be
\int [dk_{n}]: \gvk \gvq : e^{\sum _{n>0} k_{-n}q_{n}n +k_{0}.q_{0}}
e^{1/2\sum_{n,m= -\infty}^{\infty}k_{n}[ \theta _{nm} -
\lambda ^{-1}_{nm}] k_{m}}
\ee
which becomes (on doing the $k$ integration)
\br
&
:e^{1/2\sum _{n,m>0}Y_{n}
(\theta - \lambda ^{-1})^{-1}_{n,m}Y_{m}
 -1/2\sum_{n,m\geq 0}nq_{n}
(\theta - \lambda ^{-1})^{-1}_{-n,-m}
mq_{m} - i \sum _{n\geq 0,m>0}nq_{n}
(\theta - \lambda ^{-1})^{-1}_{-n,m}Y_{m}}:  \nonumber \\
&
\{ Det[\tli ] \} ^{D/2}
\er
In (2.16) to save space we have assumed implicitly that $nq_{n}$ is to be
replaced by $q_{0}$ when $n=0$.
    If one defines the column vector
\be
{\bf Y} \equiv \left(\begin{array}{r} Y_{n} \\ -inq_{n} \end{array}\right)
\equiv \left( \begin{array}{l} Y_{n} \\ Y_{-n} \end{array} \right)
\ee
with the understanding that $n=0$ is counted as a negative index (i.e.
there is no $Y_{+0}=X$ in (2.17)),then (2.15) can be written compactly as
\be
:e^{1/2 {\bf Y}^{T} \tli {\bf Y} } \gvq : [Det (1- \la \theta )]^{-D/2}
\ee
We have dropped an overall factor of $Det \la ^{D/2}$. Using
$\tli = -\left( \frac{1}{(1-\lambda  \theta )} \right) \lambda
\equiv -\Pl $
we finally get:
\be
e^{ \la _{n} L_{-n}}: \gvq : =:e^{-1/2{\bf Y}^{T} \Pl {\bf Y}} \gvq :
[Det( 1- \la \theta )]^{-D/2}
\ee
This is a simple looking result.  Furthermore it required very little
effort using the loop variable formalism. Although as pointed out
earlier it is not quite correct, the correct result is very similar
to this and is also easy to derive and we will do this in the next
section. (We need a more precise treatment of the normal ordering.)

Let us first see why the present result cannot be correct.
This becomes evident when one checks the
group composition property by calculating $ e^{\mu _{n} L_{-n}} e^
{\lambda _{n}L_{-n}} \gvq $. This requires us to calculate the action of
\emn\ on (2.19).  This is not very difficult if one realizes that written
in the form (2.15), (2.19) is just another generalized vertex operator
:$e^{i(k_{n} +q_{n})Y_{n}}$: , multiplied by some Y-independent factors.
  So we first calculate the action of \emn\ on
  :$e^{i(k_{n} + q_{n} )Y{n}}:$
and then perform the $k $ integration.  And of course for \emn\
we can again use a first order representation:
\be
\emn = \int {\cal D} p(t) : \lpp : e^{1/2 p_{n} [ \tm ]_{nm} p_{m}}
\ee
Acting on (2.15) gives
\begin{eqnarray}
\int {\cal D}p(t) :\lpp ::e^{i\sum _{n>0}(k_{n}+q_{n})Y_{n} +q_{0}Y_{0}}:
e^{k_{-n}.q_{n}n + k_{0}q_{0}} \nonumber \\
e^{+1/2k_{n}[\tl ]_{nm}k_{m} +1/2p_{n}[\tm ]_{nm}p_{m}}
\end{eqnarray}
As before $nq_{n}$ becomes $q_{0}$ for $n=0$.  We calculate the OPE using
(2.13) again to get
\br
\int {\cal D}p(t) :\lpp   e^{i\sum _{n>0}(k_{n}+q_{n})Y_{n} +q_{0}Y_{0}}:
e^{p_{-n}.(k_{n}+ q_{n})n +  k_{-n}.q_{n}n  + p_{0}.q_{0} +k_{0}.q_{0}}
\nonumber
 \\
e^{+1/2k_{n}[\tl ]_{nm}k_{m} +1/2p_{n}[\tm ]_{nm}p_{m}}
\er
We can Taylor expand $\partial X(z+t)$ as before to get (reinserting
$\int [dk_{n}]$):
\br
\int[dk_{n}] \int d[p_{n}]
:  e^{i\sum _{n>0}(p_{n} + k_{n}+q_{n})Y_{n} +iq_{0}Y_{0}}:
e^{p_{-n}.(k_{n}+ q_{n})n +  k_{-n}.q_{n}n  + p_{0}.q_{0} +k_{0}q_{0}}
\nonumber \\
e^{+1/2k_{n}[\tl ]_{nm}k_{m} +1/2p_{n}[\tm ]_{nm}p_{m}}
\er
This can be written compactly if we define the column vectors
\be
{\cal P} = \left( \begin{array}{l}p_{+n} \\ p_{-n} \\ k_{+n} \\ k_{-n}
\end{array} \right)      {\cal Y} = \left( \begin{array}{l} Y_{+n} \\
Y_{-n} \\ Y_{+n} \\Y_{-n} \end{array} \right) = \left( \begin{array}{c}
Y_{+n} \\ -inq_{n} \\ Y_{+n} \\ -inq_{n} \end{array} \right)
=\left(\begin{array}{c} {\bf Y} \\ {\bf Y} \end{array} \right)
\ee
and the matrix
\be
N = \left( \begin{array}{ll} N_{+n,+m} & N_{+n,-m} \\ N_{-n,+m} & N_{-n,-m}
\end{array} \right)  = \left( \begin{array}{cc} 0 & 0 \\ n\delta _{n,m} & 0
\end{array} \right)
\ee

(2.23) becomes
\be
\int [d {\cal P}] e^{i{\cal P}^{T} {\cal Y} } e^{1/2 {\cal P} ^{T}
\left( \begin{array}{cc} \tm & N \\ \NT & \tl \end{array} \right)
{\cal P} }
\gvq
\ee
[Note that since $Y_{+0} \equiv 0$, we can also assume $p_{+0} \equiv
k_{+0} \equiv 0$ and $p_{-0} \equiv p_{0} ; k_{-0} \equiv k_{0}$]
    Let us define

\be
K = \left( \begin{array}{cc} \tm & 0 \\ 0 & \tl \end{array} \right) \nonumber
\\  and  M = \left( \begin{array}{cc} 0 & N \\ \NT & 0\end{array} \right)
\ee
Then
\be
K^{-1} = \left( \begin{array}{cc} \tmi & 0 \\
0 & \tli \end{array} \right)
 = \left( \begin{array}{cc}-\Pm  & 0 \\
0 &-\Pl  \end{array} \right) \equiv -P
\ee
Performing the Gaussian integrals we get
\be
: e^{-1/2 {\cal Y}^{T} (K+M)^{-1}{\cal Y}} \gvq : Det [K+M] ^{-D/2}
\ee
Using
\be
(K+M)^{-1}= \left( \frac{1}{1-PM} \right) P = P + PMP + PMPMP + \cdots
\ee
this becomes
\br
: e^{-1/2 {\bf Y}^{T} ( \Pm + \Pl + \Pm \N \Pl + \Pl \NT \Pm
+ \Pm \N \Pl \NT \Pm + \Pl \NT \Pm \N \Pl + \cdots ) {\bf Y}} \nonumber \\
\gvq :
Det [K+M]^{-D/2}
\er
    If we set $\mu = \la $ in (2.31) we should just end up with
$P^{2\lambda}$. But one can check that \Pl  \  as
defined in (2.18) and (2.19) does not satisfy
this.  This shows that this result is not quite correct.  It turns out
with some modification of $\theta $ the final result (see (1.5))  has
the same
general form  and is not much harder to derive.
Except for this modification of the form of $\theta$ (and therefore
of $\Pl$) the calculation proceeds {\em exactly as above}. The
argument that leads to the correct expression for $\theta$ can be
motivated by looking at (2.31).
Note that $\theta $ in (2.18) is $1/2(N+\NT )$.  From (2.31) we see
that $N$ and \NT always occur either as $\Pm N \Pl$ or as \Pl \NT \Pm\,
i.e. they keep track of the order of \Pm vis a vis \Pl. We can generalize
this rule as follows.
We imagine
rewriting \eln\ as
$e^{ \frac {\la ^{1} _{n} L_{-n}}{N}}
e^{ \frac {\la ^{2} _{n} L_{-n}}{N}}  \cdots
e^{ \frac {\la ^{N} _{n} L_{-n}}{N}}$ for some large number $N$
(not to be confused with the matrix $N$).  In the end we set
$\la ^{1} = \la ^{2} = \cdots = \la ^{N} $
We would expect based on the above that in a particular sequence of
$ \la N \la N \la \NT \cdots \la$
whether we should have an \N or \NT is dependent on which factor of
$e^{\frac{\la ^{i}_{n} L_{-n}}{N}}$ the particular \la (adjacent to that N or
\NT\ ) came from.  Thus the rule is to have $\la ^{i} N \la ^{j}$ if $i<j$
and $ \la ^{i} \NT \la ^{j} $ if $ i>j$.  Thus we have a counting problem:
The coefficient
of a term with a particular sequence of $N$'s and \NT\ 's is equal to the
number of ways we can arrange the $\la ^{i}$'s (in accordance with the
above rule) in order to get this particular sequence of $N$'s
and \NT\ 's. Remember that
in the end the \la 's are all set equal to each other.  Implementing this rule
modifies \Pl\ defined in (2.19) to the one defined in (1.5) with a more
elaborate $\theta $ matrix.  We elaborate on this in the next section
and prove that (1.5) is the right result.

\newpage
\section{Proof}
\setcounter{equation}{0}
We first outline the steps involved in the proof and then fill in the
details:

{\bf Step1}:  To linear order in \la \ the expression (1.4) and (1.5)
for \eln\ is correct.

{\bf Step 2}: To bilinear order in \la \ $\mu$ , expression (2.31) for
\eln \ \emn \
is correct and the Virasoro Algebra is satisfied.

{\bf Step 3}: To trilinear order in  \la \ , $ \mu ,\rho $ the expression for
\eln \ \emn \ \ern \ :\gvq \ : is
\[ e^{-1/2{\bf Y}^{T} ( \Prh + \Pm + \Pl + \Pm \N \Pl + \Pl \NT \Pm +
\Prh \N \Pm
+ \Prh \N \Pl + \Pm \NT \Prh + \Pl \NT \Prh +} \]
\[ \Prh \N \Pm N \Pl +
\Pm \N \Pl \NT \Prh + \Pl \NT \Prh \N \Pm + \Pm \NT \Prh \N \Pl + \]
\[   \Prh \N \Pl \NT \Pm +
  \Pl \NT \Pm \NT \Prh + \cdots ) {\bf Y} \]
\be
\gvq
Det [K+M]^{-D/2}
\ee
Here $K=P^{-1}$.
where
\be
P= \left( \begin{array}{ccc}
\Prh & 0  &  0 \\
0  & \Pm &  0  \\
0  & 0   &  \Pl
\end{array} \right)
\ee
and
\be
  M = \left( \begin{array}{ccc} 0   &  N  &  N  \\
                                  \NT &  0  &  N  \\
                                  \NT & \NT &  0
\end{array} \right).
\ee

The rule for writing down the terms is (as mentioned in the last section)
that between any two \Prh \ , \Pm \ , \Pl  we have an $N$ if the sequence is
part of $\rho \, \mu \, \la \,$ and \NT if the order is interchanged.

{\bf Step 4}:To $p$'th order, if we have
$e^{\lambda ^{p}_{n}L_{-n}}e^{\lambda ^{p-1}_{n}L_{-n}} ...
e^{\lambda ^{i}_{n}L_{-n}} ... e^{\lambda ^{2}_{n}L_{-n}}
e^{\lambda ^{1}_{n}L_{-n}} \gvq $
the $p$-linear term in $\la ^{1} \la ^{2} .... \la ^{p}$ consists
of a sequence of $\la ^{i}$'s and \N \ or \NT \ with an \N \ between
$\la ^{i} $ and $\la  ^{j}$ in the form $\la ^{i} N \la ^{j}$ if $i>j$
and \NT \ (i.e. $\la ^{i} \NT \la^{j}$ ) if $i<j$.  This gives the action of
$( \la ^{p}_{n} L_{-n})(\la^{p-1} _{n}L_{-n})...
(\la^{1}_{n}L_{-n}) \gvq \mid 0>
$.

{\bf Step 5}:If we set $ \la ^{1}_{n} = \la ^{2}_{n} =...= \la ^{p}_{n}$
then step 4 gives us the $p$th order part of \eln \gvq \ upto a factor of
$1/p!$, i.e. it gives $ ( \la _{n} L_{-n} ) ^{p} \gvq $.

{\bf Step 6}: Using steps 4 and 5 we get a prescription for writing
down the
$p$th order term of  \eln \gvq \ . Take all possible orderings
$ \la ^{1} ....\la ^{p} $ placing \N \ 's and \NT \ 's as described
in step 4. When we set $ \la ^{1} = \la ^{2} = .... = \la ^{p} $
all terms with a given
sequence of \N \ 's and \NT \ 's become identical and add together.
Thus the  {\em number} of ways we can order the $\la ^{i}$ 's
to give that particular
sequence (of \N \ 's and \NT \ 's ) , divided by $p!$ ,
is the coefficient of of that particular $p$ th order term
in \eln \gvq \ .

{\bf Step 7}: The expression (1.5) implements this rule to any order
in \la.

    Let us fill in the details now:

{\bf Step 1}:
\br
& :1/2 \partial _{z} X(z+t) \partial _{z} X(z+t) : : \gvq : = \nonumber
\\ & :1/2 \partial _{z} X(z+t) \partial _{z} X(z+t)  \gvq : \, \nonumber
\\
& +
:[ \sum _{n>0} (q_{n} \partial _{z} X(z+t) \frac{n}{t^{n+1}}) \,
\nonumber \\ & +
q_{0} \partial _{z} X(z+t) \frac{1}{t} ] \gvq :  \nonumber \\ &
+ \sum _{n,m} \frac{nq_{n}.mq_{m}}{t^{n+m+2}} .
\er
gives the O.P.E. with the energy momentum tensor.  Using
\be
\partial _{z} X(z+t) = \partial _{z}X +t \partial ^{2} X + t^{2}
\partial ^{3}X /2! + \cdots + t^{n-1} \partial ^{n} X /(n-1)! +
\cdots = \sum _{n} t^{n-1} Y_{n}
\ee

we get
\be
\partial _{z} X(z+t) \partial _{z} X(z+t) = \sum _{n,m} t^{n+m-2} Y_{n}.Y_{m}
\ee
\be
q_{n} \partial _{z} X(z+t) nt^{-n-1} = \sum _{m} nq_{n}.Y_{m} t^{m-n-2}
\ee
Multiplying (3.4) by $\int _{c} dt \la (t) = \int _{c} dt \la _{p}
t ^{-p+1}$

we get for the RHS:
\be
\sum _{n=1}^{p} \la _{p} Y_{n}.Y_{p-n}  + \sum _{n \geq 0 }
\la _{p}nq_{n}
.Y_{p+n} + \sum _{n=1} ^{p} \la _{-p} nq_{n}.(p-n)q_{p-n}
+ \sum _{n \geq p} \la _{-p} nq_{n}.Y_{n-p}
\ee
(As always $nq_{n}$ is replaced by $q_{0}$ when $n=0$).  Since
$\la _{m,n} = \la _{m+n}$ this term is exactly the exponent of (1.4) with
$\Pl = \la$.  Thus the term $e^{1/2 {\bf Y}^{T} \Pl {\bf Y}}$ is
correct to linear
order in \la \ . This concludes step 1.
At this juncture we should point out one fact.  At higher
orders in \la \ there are two kinds of terms.  One of them is of the form
${\bf Y}^{T} \la \N \la ... \la {\bf Y}$.  These are the "non-trivial"
terms we are concerned about.  Then there are terms of the form $({\bf Y}
^{T} \la {\bf Y})^{m}/m! $ that come from expanding the exponential
in (1.4).
This term comes from the repeated (m-fold) action of $\la _{n} L_{-n}$
on \gvq \ directly (i.e. not from commutation of the $L_{n}$'s amongst
themselves).  In the language of the operator product expansion these are
terms that {\em do not} involve contraction of $X$'s from different
$T_{zz}$'s
with each other.  This is a "trivial" higher order \la \ -dependence
(i.e.
trivial in that it does not require any effort to prove that these terms
are present in the right way in (1.4)).

{\bf Step 2}: This step is obvious since the action of \eln \ results in
a generalized vertex operator just like the one we started with, as
can be seen from (2.15).  If this is correct to O(\la \ ), and the
subsequent action of \emn \ is correct to O( $\mu $) then the composition
\emn \eln \gvq \ must be correct to O($\mu $\la ).  For completeness
we present an explicit verification of the Virasoro algebra with central
charge, in the Appendix.  This concludes step 2.

{\bf Step 3}: The end point of step 2 is (2.31).  (2.31) is equivalent to
(2.23) which is a generalized vertex operator.  One can act on this
operator by \ern \ in exactly the same way as in in the preceding two
steps.  If we repeat (with three parameters) the steps leading from
(2.23) to (2.31) we end up with (3.1).  The algebra is extremely
straightforward so we will not repeat it here.  By the same logic that
was used in step 2, this procedure is guaranteed to give the right
answer to (non-trivial) order $\rho \mu \la $.  This concludes step 3.

{\bf Step 4:} This is a $p$-parameter generalization of step 3 and there
is nothing
new here.  This concludes step 4.

Step 5 and Step 6 are obvious.

{\bf Step 7:} Consider a typical term in (1.5):
\be
\int _{0}^{1} d \s \int _{0}^{1} d \st  \cdots \int _{0}^{1}
d \sigma _{m}
\la (\s ) \th (\s ,\st ) \la ( \st ) \th (\st , \sth ) \la ( \sth )
\cdots \th ( \sigma _{p-1} , \sigma _{p} ) \la ( \sigma _{p} ).
\ee
    Pick a particular sequence of \N \ and \NT \ .  The
    $\theta $- functions
enforce that $\st   > \s $ , $\st  > \sth $ etc.
\be
\int _{0}^{1} d \s \int _{0}^{1} d \st  \cdots \int _{0}^{1}
d \sigma _{m}
\la (\s ) \N \theta  (\s -\st ) \la ( \st ) \NT \theta  (\sth - \st )
\la ( \sth )
\cdots \theta  ( \sigma _{p-1} - \sigma _{p} ) \la ( \sigma _{p} ).
\ee

The region $0 \leq \s , \st , \cdots \sigma _{p} \leq 1 $ can be
decomposed into $p!$ regions each with a particular ordering of the
$\sigma $'s:
\be
0 \leq \sigma _{i_{1}} < \sigma _{i_{2}} < \cdots < \sigma _{i_{p}}
\leq 1 .
\ee
Thus (3.10) can be decomposed into $p!$ integrals.  Each of the
$p!$ terms can only give one of two possible values.  The
$\theta $-functions either vanish in the entire range of one of
these integrals
or equal one in the entire range., for if you have $\theta ( \sigma _{i}
- \sigma _{j} )$ the integration region either has $\sigma _{i} >
\sigma _{j} $ over the entire range, in which case $\theta ( \sigma _{i}
-\sigma _{j} ) =1$ or it has $\sigma _{i} < \sigma _{j}$ over the entire
range in which case $\theta ( \sigma _{i} - \sigma _{j} ) =0$. If any
of the $\theta $-functions are zero the integral vanishes.  If all the
$\theta $-functions are equal to one we get
(Note that  \la \ \,  does not depend on $\sigma $)
\be
\int _{0}^{1} d \sigma _{i_{1}}
 \int _{0}^{\sigma _{i_{1}}} d \sigma _{i_{2}}    \cdots
\int _{0} ^{\sigma _{i_{p}}}
d \sigma _{i_{p}}
(\la \N  \la  \NT \la
\cdots  )
= \frac{1}{p!}( \la \N \la  \NT \la \cdots )
\ee
Thus whenever the ordering of the $\sigma $'s is such that it gives
the particular sequence of \N \ 's and \NT \ 's we get a factor of $1/p!$.
The full integral (3.10) thus tries all possible ($p!$) orderings of
$\sigma _{i}$ 's and it therefore counts the number of ways it can be done.
This concludes Step 7 and thus the proof of (1.4).
\newpage
\section{Group Composition}
\setcounter{equation}{0}
    We briefly discuss the group composition property and give an
example of an explicit calculation for concreteness.  Equation (2.31)
contains the basic composition rule:  Write
\be
\emn \eln = \ern e^{\rho_{c}(\mu ,\lambda)}
\ee
$\rho_{c}$ refers to an overall normalization that depends on $\mu ,
 \lambda $.  Comparing (1.4), (2.31) and (4.1) we see that
\be
P^{\rho ( \mu , \lambda )}= \Pm + \Pl + \Pm \N \Pl + \Pl \NT \Pm + \cdots
; \rho _{c} = D/2 Trln [K+M]
\ee
This defines $\rho$ implicitly since (1.5) gives
\be
\Prh = \int d \s \left( \frac{1}{1- \rho \th } \right) _{ \s , \st }
\rho _{ \st }
\ee
Note that if we set $\mu = \la $ we get
\be
P^{2 \la } = 2 \Pl + \Pl (\N + \NT ) \Pl + \cdots
\ee
(4.4) is a non linear equation for \Pl \ .  It can be used to determine
\Pl recursively.  If we set
\be
\Pl = \la + a \la \N \la + b \la \NT \la + c \la \N \la \N \la + \cdots
\ee
one can solve recursively for the coefficients $a,b,c,...$
by substituting
(4.5) into (4.4).  One can check that the result reproduces (1.5).
In fact
one can prove that (1.5) satisfies this equation.  We will not do this
here
but the outline of the proof is as follows: Define $\la ( \sigma ) :
0< \sigma < 2 $ so that
$\la ( \sigma ) \begin{array}{ccc}
 = & \la  &  0< \sigma < 1 \\
                = & \mu  &  1< \sigma < 2  \end{array}$
One then shows that (4.2) can be written in the form (4.3) except that
the
range of $\sigma $ is from 0 to 2.  Doubling the range of $\sigma$
is equivalent to scaling $\la \rightarrow 2 \la $.  This proves that
(1.5) satisfies (4.4).  This is an alternate proof of the correctness
of (1.5).  Finally one can also show that setting $\mu = - \la$
gives $\Prh =0$.  To lowest order this is obvious since $\Pm = \mu $,
however it is not at all obvious a priori that this is satisfied by
the full expression (4.2).  We have checked this but will not reproduce
the argument here.

We conclude this section with an example.
\subsection{Example}
\be
e^{( \la _{2}L_{-2} + \la _{-2} L_{2} )}e^{ik_{0}.Y} |0>
\ee
The exact answer of course is (1.4). We can check this order by order.
Some of the non-zero elements of the matrix $\la _{n,m} $are :
\[
\la _{1,1} = \la _{0,2} = \la _{2,0} = \la _{-1,3} = \la _{3,-1}
=\la _{4,-2}= \la_{-2,4} = \la _{2}    \]
\be
\la _{-1,-1} = \la_{0,-2} = \la_{-2,0} = \la _{1,-3} = \la _{-3,1}
\la _{-4,2} = \la _{2,-4} = \la _{-2}
\ee
\be
\Pl = \la + 1/2 \la ( \N + \NT ) \la + O( \la ^{3} )
\ee
\be
{\bf Y}^{T} \la {\bf Y} = ( Y_{n} , -ink_{n} ) \left( \begin{array}{cc}
\la _{+n,+m} & \la _{+n,-m} \\ \la _{-n,+m} & \la _{-n,-m} \end{array}
\right) \left( \begin{array}{c}  Y_{m} \\ -imk_{m} \end{array} \right)
\ee

In this example only $k_{0}$ is non zero.  This simplifies (4.9)
considerably: All the '$-m$' indices in the matrix can be set to zero.
We remind the reader of our convention that '$0$' belongs to the
negative index set and '$+m$' is assumed to be a positive integer.  This
fact, along with (4.7), quickly reduces (4.9) to the equation:
\be
-1/2 {\bf Y}^{T} \la {\bf Y} = - \la _{2} ( Y_{1}.Y_{1} + 2 Y_{2} (-ik_{0}))
\ee
This is just $L_{-2} e^{ik_{0}X}|0> $.
At the next order in $ \la $ we have to calculate
\br
& {\bf Y} ^{T} ( 1/2 \la ( \N + \NT ) \la ) {\bf Y} =  \\ &
(Y_{n} -ik_{0} )
\left( \begin{array}{cc} \la _{+n,+m} & \la _{+n,-m} \\ \la _{-n,+m} &
\la _{-n,-m} \end{array} \right) \left( \begin{array}{cc} 0 & m \\
 m & 0 \end{array} \right) \left( \begin{array} {cc} \la _{+m,+p} &

\la _{+m,-p} \\ \la _{-m,+p} & \la _{-m,-p} \end{array} \right)
\left( \begin{array}{c} Y_{p} \\ -ik_{0} \end{array} \right) \nonumber
\er
As before the '$-p$' index is forced to be '$0$', as is the '$-n$' index.
Once again using (4.7) we get finally
\br
& -1/4 {\bf Y}^{T} \la ( \N + \NT ) \la {\bf Y} +
1/8 ( {\bf Y}^{T} \la {\bf Y} ) ^{2}  \nonumber \\ &
= -1/4 \la _{2} ^{2} (2Y_{1}.Y_{3} - 4ik_{0}.Y_{4}) +
1/4 \la _{2} \la _{-2}
4 k_{0} ^{2} + \nonumber \\ &
\la ^{2} _{2} /2 (Y_{1}.Y_{1} + 2 Y_{2}(-ik_{0}))^{2}
\er
where we have added the 'trivial' second order contribution to (4.10).
One also has to add a contribution from the determinant at this order.
\be
Det[ \delta _{ \s \st } - \la _{ \s } \th _{ \s \st } ] ^{-D/2}
= e ^{-D/2 Tr ln [ 1- \la \th ]} = e ^{D/2 ( Tr \la \th + 1/2 Tr ( \la \th
\la \th ) ... )}
\ee
The trace includes one over the $\sigma $ - index as well as the $n$- index.
\be
Tr \la \th = Tr \int d \s \int d \st [ \la \N \theta ( \s - \st ) +
\la \NT \theta ( \st - \s ) ] \delta ( \s - \st ) = 0
\ee
since $Tr \la \N =0$.
At the next order we have
\[
Tr \int d \s \int d \st \la[ \N \theta ( \s - \st ) + \NT \theta ( \st - \s )]
\la [ \N \theta ( \st - \s ) + \NT \theta ( \s - \st ) ]
 \]
\[
= 2 tr \int d \s \int d \st \la \N \la \NT \theta ( \s - \st )
\]
\[
=1/2 2 Tr \la \N \la \NT = \la _{2} \la _{-2}
\]
Thus the contribution of the determinant is
\be
D/4 \la _{2} \la _{-2}
\ee
Since
\[ L_{2} L_{-2} e^{ik_{0}X} |0> = [L_{2},L_{-2}]e^{ik_{0}X}|0> \]
\[(4L_{0} + D/2 ) e^{ik_{0}X} |0> = 2k_{0}^{2} + D/2 ) e^{ik_{0} X}|0> \]
we see that (4.12 ) and (4.15) do indeed give the right answer.
\newpage
\section{Conclusions}
\setcounter{equation}{0}
Let us summarize the results.  We have given an expression in closed
form for the action of an element of the Virasoro group on generalized
vertex operators \gvk \ in the form
$\eln \gvk |o>$.  The result is given in equation (1.4).  The main idea
was to
use a first order formalism for the energy momentum tensor which recasts
the group element in the form of a loop variable \lp  \ .  This makes the
calculation very easy.  The crucial issue that has to be addressed is that
of regularizing the loop variable or equivalently taking into account
contractions of the $X$ field inside the loop variable. We sidestepped
this question by building up a finite group element as a product
of infinitesimal group elements, each represented as a loop variable
for which we do not have self contractions (since we only keep the linear
piece of each infinitesimal element)
We were able to derive the expression (1.4) in this manner.

The naive approach
described in section II does not give the (right) answer (1.4) although
it gives something very similar looking.
It would give some insight into the formalism of loop variables if we
could identify a procedure that modifies the naive approach to
take into account the self
interactions of the
 $X$ fields in a loop variable in a self consistent way and leads
to (1.4) directly.  It turns out that this is not very difficult.
The basic idea is to thicken the loop into a band or an annulus
as shown in
fig 2.  Then we repeat the method of section 2 but now applied to a
superposition of all the loops making a band ,i.e. to variables of the
form $\int d \sigma \int dt k(t, \sigma ) \partial X(z+t) $. $\sigma $ is
assumed to vary from 0 to 1 and parametrizes the different loops
in the annulus.
Thus we let $ \sigma =0$ correspond to the inner boundary of
the annulus
and $\sigma =1$ correspond to the outer boundary. Then (2.5)
is modified to
\br
& \bp = \nonumber \\ &
: \bp : e^{\int dt \int dt'\int d \sigma d \sigma '
k(t, \sigma ) k(t' , \sigma ') < \partial X(z+t) \partial X(Z+t') > }
\er

For $ \sigma > \sigma ' $ we take $t$ to be the outer loop and
$t'$ to be the inner loop and vice versa when $ \sigma < \sigma ' $.
The inner loop
can be shrunk to a point since there are no singularities and we
can expand
$\partial X(z+t) $ in a Taylor series with only positive powers of $t$.
The result is
\br
& \bp = \nonumber \\ &
: \bp : e^{ \int d \sigma \int d \sigma '
\sum _{n \geq 0}(k_{n}( \sigma )
k_{-n} ( \sigma ')[n \theta ( \sigma - \sigma ')] +
k_{-n}( \sigma ) k_{n}
( \sigma ') [n \theta ( \sigma ' - \sigma )])}
\er
This can be written as
\be
e^{\int d \sigma \int d \sigma ' k_{n}(\sigma ) k_{m} (\sigma ')
[\N \theta ( \sigma - \sigma ' ) +
\NT \theta ( \sigma ' - \sigma )]_{nm}}
\ee
The expression in square brackets is $\th $ of eq.(1.3).  If we
include the wave function as before, the complete first order form for
the group element is
\be
\int {\cal D} k(t, \sigma ) : \bp : e^{\int d \sigma \int d \sigma '
k_{n} ( \sigma ) k_{m} ( \sigma ')
[ \th _{ \s \st } - \delta _{ \s \st } \la
( \s ) ]_{nm}}
\ee
Taking the OP of this loop (or 'band') variable with a generalized
vertex operator
\gvq \ and doing the $k$-integration gives the result (1.4).  Intuitively
the group element \eln \ is being written as a product of infinitesimal
elements :
\be
\underbrace{e^{ \frac{\la _{n} L_{-n}}{N}}e^{ \frac{\la _{n}
L_{-n}}{N}} \cdots
e^{ \frac{\la _{n} L_{-n}}{N}}}_{N - factors} ;
 N \rightarrow \infty
\ee
Each loop in the annulus represents one of these factors.  The
contractions between the $X$-fields of different loops represent the
nontrivial commutation between the $L_{n}$'s coming from different
factors.

We can also compare the result obtained here with that in \cite{XVIII}.
The result obtained
there was that the action of a sequence of $L_{n}$'s on a vertex
operator \gvk \ can be obtained by starting with
\be
e^{-1/2 Y     q \frac{1}{1- \theta q} Y + ik \frac{1}{1- \theta q} Y
-1/2 k\frac{1}{1- \theta q} \theta k } Det(1- \theta q )^{-D/2}
\ee
where $k_{n} , q_{nm}$ and $\theta _{nm} $ were variables
(defined in \cite{XVIII}) with
simple transformation properties under the Virasoro algebra.  Comparing
(5.6) with (1.4) and (2.15) shows that they are very similar looking.  It
must be true then that $q_{nm} , \theta _{nm}$ are variables that
parametrize the group so that they should correspond more or less to
$\la _{+n,+m} , \la _{-n,-m} and \la _{+n,-m}$.  While we know the
transformation properties of $ q, \theta $  we do not know how they
parametrize
the group.  On the other hand in the case of the  \la \ variables of this
paper , we know how they parametrize the group but we have not worked
out the transformation properties.  Thus the results obtained here
complement
those of \cite{XVIII}
To make a precise comparison one would have to work
out the transformations of the \la \ 's and compare with those of the
$q , \theta $ variables. This is in progress.

Finally, on a more speculative level we might hope that some of
the techniques
used here might be useful for string interactions.  The action of
 a group element on a vertex operator is related to the kinetic term in
string field theory. Thus if we can write the kinetic term as an
OP of two
loop variables as done here we might hope for a more unified treatment of
the kinetic and interaction terms in string field theory.

\underline{Acknowledgements}: I would like to thank M. Gunaydin
for a useful discussion.
\newpage
\appendix
\section{Appendix}
\setcounter{equation}{0}
\renewcommand{\theequation}{\Alph{section}.\arabic{equation}}
We show that to linear order in \la \ and $\mu$ (2.31) is consistent
with the Virasoro Algebra.  We reproduce (2.31) for convenience

\br
& e^{-1/2 {\bf Y}^{T} \Prh {\bf Y}}\gvq Det [1- \rho \th ]^{-D/2} =
\nonumber \\ & \emn \eln \gvq = \nonumber \\ &
: e^{-1/2 {\bf  Y}^{T} ( \Pm + \Pl + \Pm \N \Pl + \Pl \NT \Pm
+ \Pm \N \Pl \NT \Pm + \Pl \NT \Pm \N \Pl + \cdots ) {\bf Y}}
\nonumber \\ &
\gvq :
Det [K+M]^{-D/2}
\er
We need to determine $\rho ( \mu , \la )$ from \Prh \ as defined above and
then calculate the commutator $ \rho (\mu , \la ) - \rho ( \la , \mu )$.
To second order in $\rho $
\be
\Prh =\rho + 1/2 \rho ( \N + \NT ) \rho + \cdots
\ee
In solving for $\rho$ to O($\mu \la $) we need to keep terms upto
second order in  \Prh \ , since $\Prh \ = \mu + \la +...$. The answer is
\be
\rho = \Prh - 1/2 \Prh ( \N + \NT ) \Prh + O ( \Prh ^{3} ) + \cdots
\ee
Substituting for \Prh \ from (2.31) we get
\[
\rho ( \mu , \la ) = \Pm + \Pl + 1/2( \Pm \N \Pl + \Pl \NT \Pm - \]
\be
  \Pm \NT \Pl - \Pl \N \Pm - \Pm ( \N + \NT ) \Pm
 - \Pl ( \N + \NT ) \Pl )
\ee
which gives
\[
\rho ( \mu , \la ) - \rho ( \la , \mu ) = \Pm \N \Pl + \Pl \NT \Pm
- \Pm \NT \Pl - \Pl \N \Pm   \]
\be
= \mu \N \la + \la \NT \mu - \mu \NT \la - \la \N \mu
\ee
to O($\mu \la $ ).

Let us choose $ \la _{a}$ and $\mu _{b}$ as the two non-vanishing
parameters.
This will give us $ [L_{a}, L_{b}]$.  From the definition of
$\la _{n,m}$ we have :
\be
\la _{a-n,n}= \la _{n, a-n} = \la _{a} ; \mu _{b-n,n} =\mu _{n,b-n} =
\mu _{b}  ; \forall n.
\ee
We then get :
\[
( \la \N \mu ) _{nq} = \sum _{m \leq 0} \la _{nm} (-m) \delta _{-m,p}
\mu _{p,q} = \]
\be
\sum _{n \geq a} \la _{n,a-n} (a-n) \mu _{n-a,b-n+a}
= - \sum _{n \geq a} \la _{a} \mu _{b} (a-n) \delta _{q,b+a-n}
\ee
\be
( \la \NT \mu )_{nq} = \sum _{n \leq a} \la _{a} \mu _{b} (a-n) \delta
_{q,b+a-n}
\ee
\be
( \la \NT \mu - \la \N \mu ) _{nq} = \sum _{n} \la _{a} \mu _{b} (a-n)
\delta _{q,b+a-n}
\ee
\be
( \mu \NT \la - \mu \N \la ) _{nq} = \sum _{n} \mu _{b} \la _{a} (a-b)
\delta _{q,b+a-n}
\ee
which gives
\be
( \rho ) _{nq} = \sum _{n} \la _{a} \mu _{b} (a-b) \delta _{q, b+a -n}
\ee
Using the definition $ \rho _{n+m} = \rho _{n,m} $ we get $ \rho _{a+b}
= \la _{a} \mu _{b} (a-b) $.  This confirms the Virasoro Algebra
except for the central charge , to which we now turn.

The central charge is $ D/2 Tr ln ( K+M) - ( \mu \leftrightarrow \la )$
which, to this order is :
\be
D/2 Tr ( \Pm \N \Pl \NT ) - ( \mu \leftrightarrow \la )
\ee
with
\[ Tr ( \Pm \N \Pl \NT ) = \sum _{m,p >0} mp \mu _{-p-m} \la _{+m+p} \].,
which gives for (A.12)
\be
\sum _{m,p >0} ( \mu _{-p-m}\la _{+m+p} - \la _{-m-p} \mu _{+m+p})mp
\ee
If we let $ \la _{a} ; \, a >1 $ and $ \mu _{-a} $ be the two non
vanishing parameters, this becomes:
\be
D/2 \sum _{m=1} ^{a-1} m(a-m) \mu _{-a} \la _{a} = \frac{(a^{3} -a)D}{12}
\mu _{-a} \la _{a}
\ee
which is the central charge.

Thus eq.(2.31) is consistent with the Virasoro Algebra with central
extension.

\end{document}